\documentclass[aps,preprint,superscriptaddress,nofootinbib]{revtex4}
\usepackage{amsmath}
\allowdisplaybreaks[1]
\usepackage{mathtools}
\usepackage{physics}
\usepackage{graphicx}
\usepackage{amssymb}
\usepackage[colorlinks,citecolor=red]{hyperref}
\usepackage{slashed}
\usepackage{bm}
\usepackage{adjustbox}

\newcommand{\beq}{\begin{equation}}
\newcommand{\eeq}{\end{equation}}

\def\GeV{{\rm GeV}}

\newcommand\snowmass{\begin{center}\rule[-0.2in]{\hsize}{0.01in}\\\rule{\hsize}{0.01in}\\
\vskip 0.1in Submitted to the  Proceedings of the US Community Study\\ 
on the Future of Particle Physics (Snowmass 2021)\\ 
\rule{\hsize}{0.01in}\\\rule[+0.2in]{\hsize}{0.01in} \end{center}}

\begin{document}

\preprint{
	{\vbox {
		\hbox{\bf MSUHEP-22-006}
}}}
\vspace*{0.2cm}

\title{Azimuthal angular correlation as a new boosted top jet substructure}

\author{Zhite Yu}
\email{yuzhite@msu.edu}
\affiliation{Department of Physics and Astronomy,
Michigan State University, East Lansing, MI 48824, USA}

\author{C.-P. Yuan}
\email{yuanch@msu.edu}
\affiliation{Department of Physics and Astronomy,
Michigan State University, East Lansing, MI 48824, USA}

\date{\today}

\begin{abstract}
When a top quark is highly boosted, the $W$ boson from its decay has a substantial linear polarization that results in a $\cos2\phi$ azimuthal angular correlation among the top decay products. We show that this correlation can be measured for hadronically decayed boosted tops, and its magnitude provides a way to measure the longitudinal polarization of top quark, which is an important probe of new physics that couples to top sector.
\end{abstract}

\snowmass

\maketitle

At a collision energy that is around one hundred times the top quark mass, the Large Hadron Collider (LHC) not only can produce top quarks, but also has a certain chance to produce them being highly boosted, with energies much greater than their mass. The boosted top quarks provide a unique ground to further test the Standard Model (SM) and to also search for new physics (NP)~\cite{Schatzel:2013wsr}. In this kinematic region, the top quark decay products are collimated and resemble in appearance to a light quark or gluon jet. Such cone signature enhances the selection efficiency of boosted top quark events with respect to the background~\citep{Abdesselam:2010pt}. In addition, the semileptonic decay mode no longer possesses special advantage over the hadronic mode, and one ought to take the latter into account to enhance the statistics. Then the boosted top quark can be readily identified as a single {\it fat} jet by some jet algorithm and become difficult to be distinguished from a QCD jet. Hence, for experimental study of boosted tops, one needs to first be able to distinguish a boosted top quark jet from a QCD jet.

\begin{figure}[htbp]
	\centering
	\includegraphics[scale=0.6]{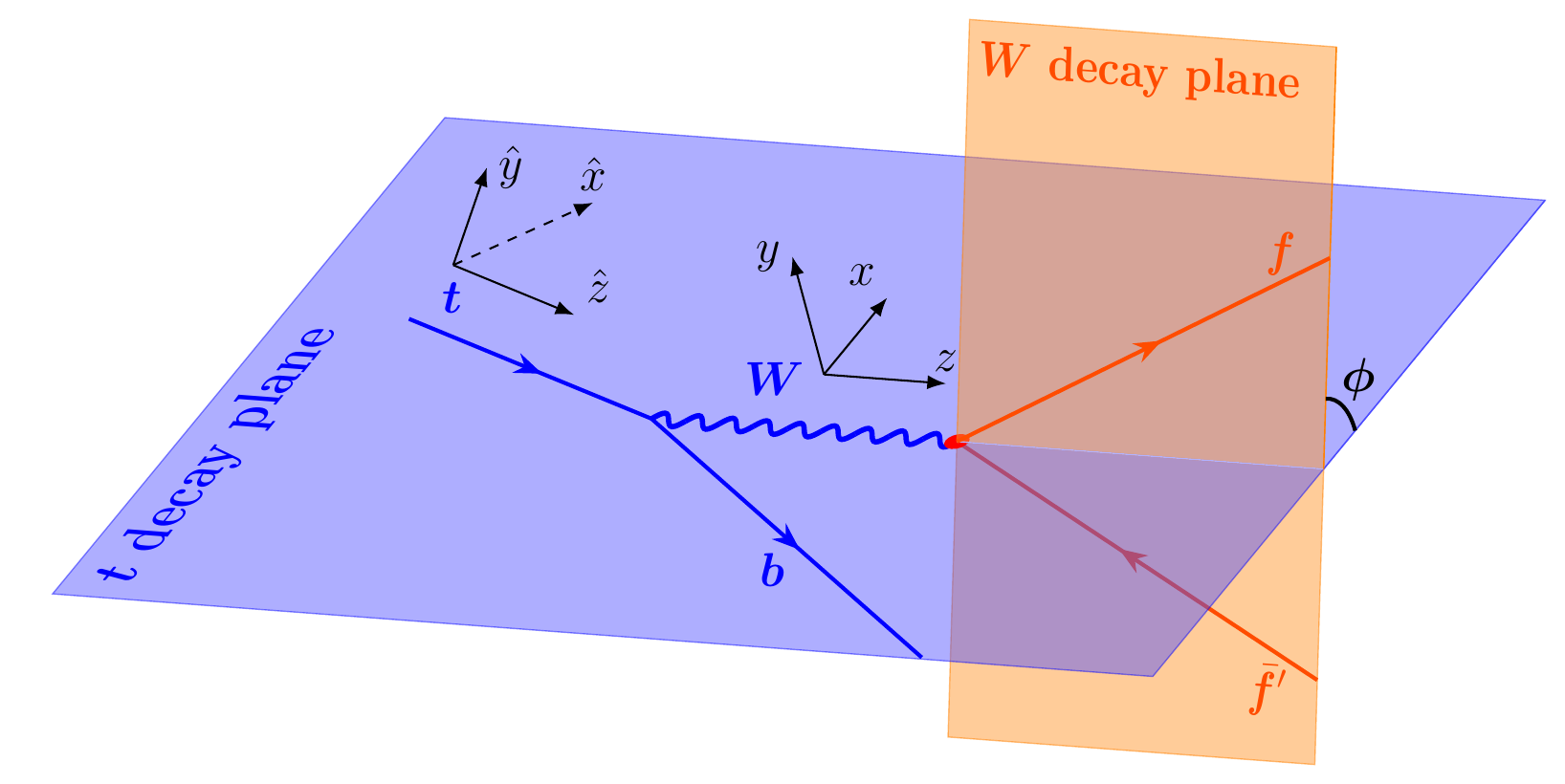}
	\caption{The two successive decay planes in $t\to bW (\to f\bar{f}')$ decay process. The coordinate systems of the top frame and the $W$ frame are shown separately. The $x$ axis of the $W$ frame lies on the $t$ decay plane, while the $\hat{x}$ axis of the $t$ frame may not.}
	\label{fig:frame}
\end{figure}

Many tagging algorithms~\cite{CMS-PAS-JME-13-007, Plehn:2010st, ATLAS:2018wis}, including machine learning technique~\cite{Bhattacharya:2020vzu}, have been proposed and applied to discriminate boosted top quark events from QCD jets, mainly by making use of the top and $W$ mass conditions and the three-subjet structure. 
In this contribution, we propose a new substructure observable of the boosted top quark jet that further explores the {\it azimuthal} angular correlation among these subjets. We point out that in the decay of a boosted top quark, the angle $\phi$ (see Fig.~\ref{fig:frame}) between the decay planes of $t\to b W$ and $W\to f \bar{f}'$ exhibits an interesting $\cos2\phi$ distribution that arises from the linear polarization of $W$ boson, which is a superposition of its $+1$ and $-1$ helicity eigenstates. Such polarization does not exist in the top rest frame but emerges as a result of $W$ helicity mixing when going to the boosted top frame. Compared to $\cos\phi$ or $\sin\phi$ correlation, the $\cos2\phi$ correlation does not require distinguishing the two subjets from the $W$ decay, so it is perfectly suitable for hadronically decayed top quarks. We propose that this azimuthal correlation be used together with other top taggers, which shall further improve the tagging efficiency.

We only focus on the hadronic decay mode of boosted top quarks.\footnote{The semileptonic decay mode requires the reconstruction of missing neutrino momentum, and contains the same $\cos2\phi$ correlation~\cite{Yu:2021zmw}.} 
The azimuthal angular correlation takes the form
\beq
\frac{\pi}{\Gamma_t}\frac{\dd{\Gamma_t}}{\dd\phi}
= 1 + \xi \cos2\phi ,
\quad
\phi \in [0, \pi),
\label{eq:phi distribution}
\eeq
where we only take $\phi \in [0, \pi)$ because we do not (or rather, we cannot, since we are considering the hadronic decay mode) distinguish the subjets from the $W$ decay. The coefficient of the $\cos2\phi$ correlation, $\xi = \xi(E_t, \lambda_t)$ is associated with the linear polarization of $W$ and depends on the top quark's energy $E_t$ and longitudinal polarization $\lambda_t$. When $E_t \gg m_t$, the dependence of $\xi$ on $E_t$ saturates very fast to the limit $E_t = \infty$, such that $E_t \gtrsim 500~\GeV$ can be considered as highly boosted, and we can approximate $\xi(E_t, \lambda_t)$ by $\xi(\infty, \lambda_t)$,
\beq
\xi(E_t, \lambda_t) \simeq \xi(\infty, \lambda_t) = 0.145(\lambda_t - 1), \quad \mbox{ as } E_t \gtrsim 500~\GeV.
\eeq
There is a similar $\cos2\phi$ azimuthal correlation in a three-pronged light QCD jet~\cite{Chen:2020adz}, which originates from a linearly polarized virtual gluon and is allowed by parity. However, the magnitude of the $\cos2\phi$ fluctuation in QCD jet is much smaller than that in the boosted top quarks, for most values of the polarization $\lambda_t$~\cite{Yu:2021zmw}. As a result, the azimuthal correlation can serve as a boosted top quark tagger against QCD jet background.

It is interesting that the parity-conserving effect, $\cos2\phi$, has a dependence on the top quark polarization $\lambda_t$. It is a result of the left-handed $Wtb$ coupling that violates parity; had we worked in a parity-conserving theory, the top polarization $\lambda_t$ would not appear, as is the case for QCD jet. On the other hand, it is the linear dependence of $\xi$ on $\lambda_t$ that enables us to measure the boosted top quark's longitudinal polarization in the hadronic decay mode. While top polarization can be easily measured in its rest frame for the semileptonic decay mode~\cite{ATLAS-CONF-2021-027, ATLAS:2013gil, Jezabek:1994qs, Brandenburg:2002xr, CMS:2019nrx, Mahlon:2010gw, Schwienhorst:2010je, Aguilar-Saavedra:2017wpl}, it is not so easy to measure in the boosted regime, and neither for the hadronic mode. In the boosted regime, one naturally chooses the helicity basis and expresses the top polarization vector as $\bm{s}_t = \left(b_1, b_2, \lambda_t \right)$ with respect to the $\hat{x}$-$\hat{y}$-$\hat{z}$ coordinate system in Fig.~\ref{fig:frame}, where $\bm{b}_T =  \left(b_1, b_2\right)$ is the transverse spin. In the top rest frame, $\bm{s}_t$ controls the angular distribution of its decay particles. Specifically, $\bm{b}_T$ is associated with the azimuthal distribution and $\lambda_t$ the polar distribution. When the top quark is boosted, the azimuthal distribution stays invariant and was suggested to be used to measure top transverse spin~\cite{Kane:1991bg}. In contrast, the polar distribution becomes too distorted to be useful for measuring $\lambda_t$. Therefore, in this contribution, we propose to use the {\it azimuthal correlation} to measure the top longitudinal polarization, which differs from the methods proposed in the literature that exploit the energy fractions of the subjets~\cite{Shelton:2008nq, Krohn:2009wm, Kitadono:2015nxf, Godbole:2019erb}.

\begin{figure}[htbp]
\centering
\includegraphics[scale=0.6]{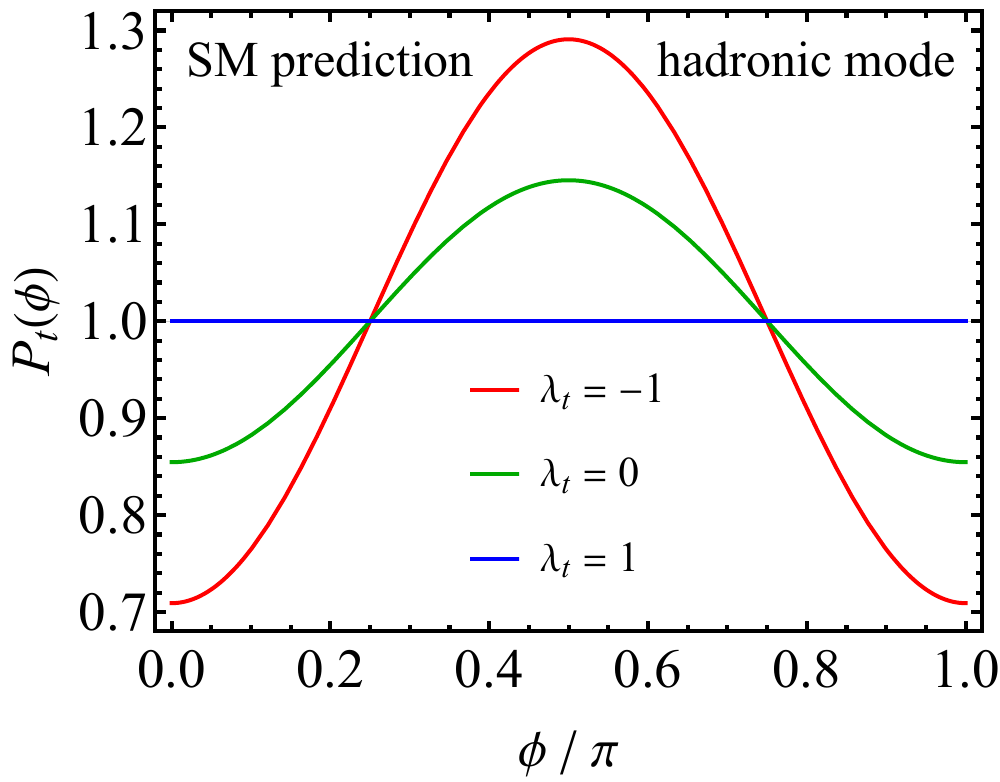}
\caption{Azimuthal angular correlation in the decay of boosted top quark for different values of top longitudinal polarization $\lambda_t$.}
\label{fig:phi distribution}
\end{figure}

The measurement of $\lambda_t$ is important to discriminate the top quark production mechanism~\cite{Berger:2011hn}, because the longitudinal polarization is highly sensitive to the chiral structure of top quark coupling to other particles. A simple example is the $W'$ model in which a heavy vector boson $W'$ can couple to top and bottom quarks in any arbitrary chiral combination~\cite{Berger:2011hn},
\beq
\mathcal{L}_{W'} = -\frac{g'}{\sqrt{2}} \bar{t} \gamma^{\mu} \left( f_L P_L + f_R P_R \right) b\, W'_{\mu} + {\rm h.c.}.
\eeq
Depending on the relative size of $f_L$ and $f_R$, the top quark can be left-handed ($\lambda_t \simeq -1$), right-handed ($\lambda_t \simeq +1$), or unpolarized ($\lambda_t \simeq 0$), corresponding to $|f_L| \gg |f_R|$, $|f_L| \ll |f_R|$, or $|f_L| \simeq |f_R|$. These three different cases lead to different azimuthal correlations, as shown in Fig.~\ref{fig:phi distribution}.

\begin{figure}[htbp]
\centering
\includegraphics[scale=0.6]{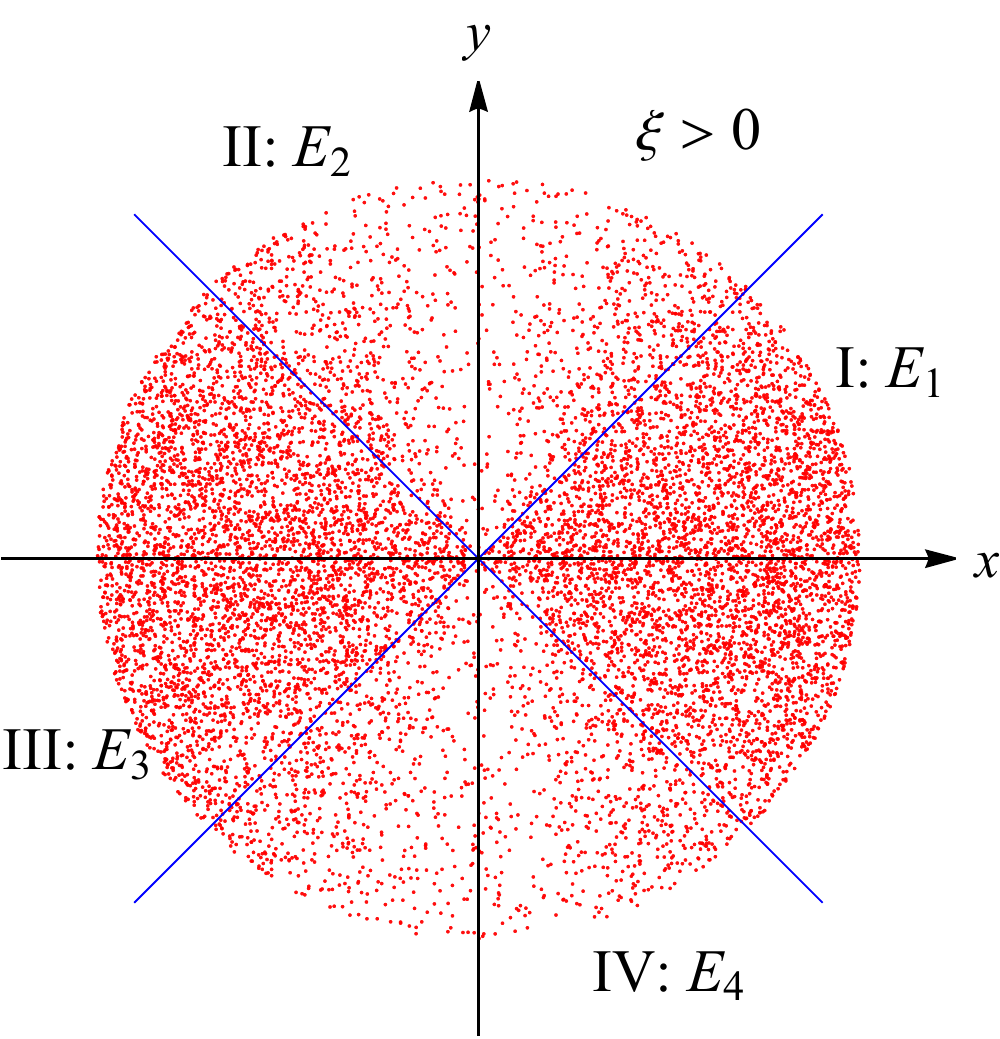}
\caption{The energy distribution of the $W$ decay products in the azimuthal plane of the $W$ frame, viewed from the $z$ direction in Fig.~\ref{fig:frame}.}
\label{fig:energy distribution}
\end{figure}

The $\cos2\phi$ correlation does not distinguish the two subjets from the $W$ decay, but only controls whether the $W$ decay plane tends to be parallel or perpendicular to the top decay plane, and this tendency carries through to the energy distribution of top decay products in the azimuthal plane. Therefore, in experimental analysis, one can extract the coefficient $\xi$ using azimuthal energy distribution, for which we propose a simple procedure that does not require the reconstruction of the $W$ decay products:
\begin{enumerate}
\item construct the top quark and its four-momentum $p_t^{\mu}$;
\item use jet substructure technique with $b$-tagging to reconstruct the $b$-subjet with its four-momentum $p_b^{\mu}$;
\item determine the $W$'s four-momentum $p_W^{\mu} = p_t^{\mu} - p_b^{\mu}$;
\item construct the $W$ frame coordinate system ($x$-$y$-$z$) as in Fig.~\ref{fig:frame}, with $z$ along $\bm{p}_W$ and $y$ along $\bm{p}_b \times \bm{p}_W$.
\item remove the particles in the $b$-subjet and determine the energy distribution of the rest of top quark jet in the transverse plane ($x$-$y$).
\end{enumerate}
With the $\cos2\phi$ modulation, the energy deposition in the transverse plane takes a form like Fig.~\ref{fig:energy distribution}, in which the azimuthal plane has been divided by $\phi = \pm\pi/4$ and $\pm3\pi/4$ into 4 quadrants, in which the cumulative energy deposits (of the whole top event ensemble) are $E_1, \cdots, E_4$, sequentially. 
Then we have
\begin{align}
\xi
= \frac{\pi}{2} \cdot \frac{(E_1 + E_3) - (E_2 + E_4) }{(E_1 + E_3) + (E_2 + E_4) },
\label{eq:exp-def}
\end{align}
where $E_{1,3}$ ($E_{2, 4}$) are associated with the region $\cos2\phi > (<) 0$.
We note that this only requires the measurement of azimuthal angles and not the identification of subjets from $W$ decay.
Also, due to its linear dependence on the energy, the definition in Eq.~\eqref{eq:exp-def} is infrared safe, not sensitive to (long-distance) soft gluon contributions.

To conclude, we proposed a novel substructure observable in the boosted top quark jet based on the azimuthal correlation between the $t\to bW$ and $W\to f\bar{f}'$ decay planes. The boosted top quark decays into a $W$ boson with a linear polarization, which results in a $\cos2\phi$ azimuthal correlation. Such linear polarization is not present in the top rest frame but only emerges under the boost as a result of mixing with other polarization parameters of the $W$ boson. We have proposed an experimental method to measure the degree of such azimuthal correlation that only requires $b$-tagging in the top quark jet. We have demonstrated that such correlation can be used to measure the longitudinal polarization of a boosted top quark for testing the SM and probing NP, and we have argued that it can also help distinguish a boosted top quark from the QCD jet background. 

\begin{acknowledgments}
This work is in part supported by the U.S.~National Science Foundation
under Grant No.~PHY-2013791. C.-P.~Yuan is also grateful for the support from the Wu-Ki Tung endowed chair in particle physics.
\end{acknowledgments}

\bibliographystyle{apsrev}
\bibliography{reference}

\end{document}